\documentclass[twocolumn]{aastex62}
\usepackage{epsf}
\usepackage{graphicx}
\usepackage{amsmath}

\begin{document}

\title{Observations and Preliminary Modeling of the Light Curves of Eclipsing Binary Systems NSVS 7322420 and NSVS 5726288}

\author{Matthew F. Knote}
\affil{Florida Institute of Technology}
\affil{Ball State University}
\author{Ronald H. Kaitchuck}
\affil{Ball State University}
\author{Robert C. Berrington}
\affil{Ball State University}
\begin{abstract}
We present new photometric observations of the $\beta$ Lyrae-type eclipsing binary systems NSVS 7322420 and NSVS 5726288. These observations represent the first multi-band photometry performed on these systems. The light curves were analyzed with PHOEBE, a front-end GUI based on the Wilson-Devinney program, to produce models to describe our observations. Our preliminary solutions indicate that NSVS 7322420 is a primary filling semi-detached system with unusual features warranting further study. These features include a pronounced O'Connell effect, a temporal variance in the light curve, and an unusual ``kink'' in the light curve around the secondary eclipse. The cause of these features is unknown, but one possibility is the transfer of mass between the component stars. Meanwhile, NSVS 5726288 is probably a typical detached system.
\end{abstract}

\section{Introduction}

Eclipsing binary systems are an important class of variable star as they allow the determination of several characteristics that are otherwise difficult or impossible to determine without resolving the stellar components. Photometric measurements allow the determination of the orbital inclination of the system as well as the shapes and radii of the component stars for totally eclipsing systems, while radial velocity measurements allow the determination of the absolute masses of the components. With both types of measurements, a nearly complete description of the system and its components can be obtained.

\begin{table}[b]
\centering
\begin{minipage}{\textwidth}
\centering
\begin{tabular}{ccccccccccc}
\hline
Star & \multicolumn{3}{c}{R.A. (J2000)} & \multicolumn{3}{c}{Dec. (J2000)} & B & V & R & I\\
Designation & h & m & s & d & m & s &  &  &  & \\
\hline
\multicolumn{9}{c}{Variable Star}\\
\hline
NSVS 7322420 & 8 & 16 & 12.90 & +26 & 41 & 13.67 & 11.675 & 10.983 & 10.538 & 10.122\\
NSVS 5726288 & 20 & 19 & 11.65 & +44 & 15 & 47.05 & 11.607 & 11.297 & 11.058 & 10.831\\
\hline 
\multicolumn{9}{c}{Comparison Star}\\
\hline
TYC 1936-1001-1  & 8 & 17 & 3.36 & +26 & 45 & 14.52 & 11.215 & 10.695 & 10.344 & 10.014\\
TYC 3163-221-1 & 20 & 18 & 51.80 & +44 & 12 & 44.81 & 11.112 & 10.781 & 10.549 & 10.328\\
\hline 
\multicolumn{9}{c}{Check Star}\\
\hline
TYC 1936-113-1 & 8 & 17 & 5.13 & +26 & 43 & 6.50 & 12.228 & 11.207 & 10.654 & 10.139\\
BD+43 3570 & 20 & 19 & 56.71 & +44 & 11 & 19.09 & 11.021 & 10.894 & 10.745 & 10.600\\
\hline
\end{tabular}
\caption{List of the designation, coordinates, and B, V, R, and I magnitudes of each variable, comparison, and check star.}
\end{minipage}
\label{tab:Systems-characteristics}
\end{table}

\begin{figure}
\begin{center}
\includegraphics[width=\columnwidth]{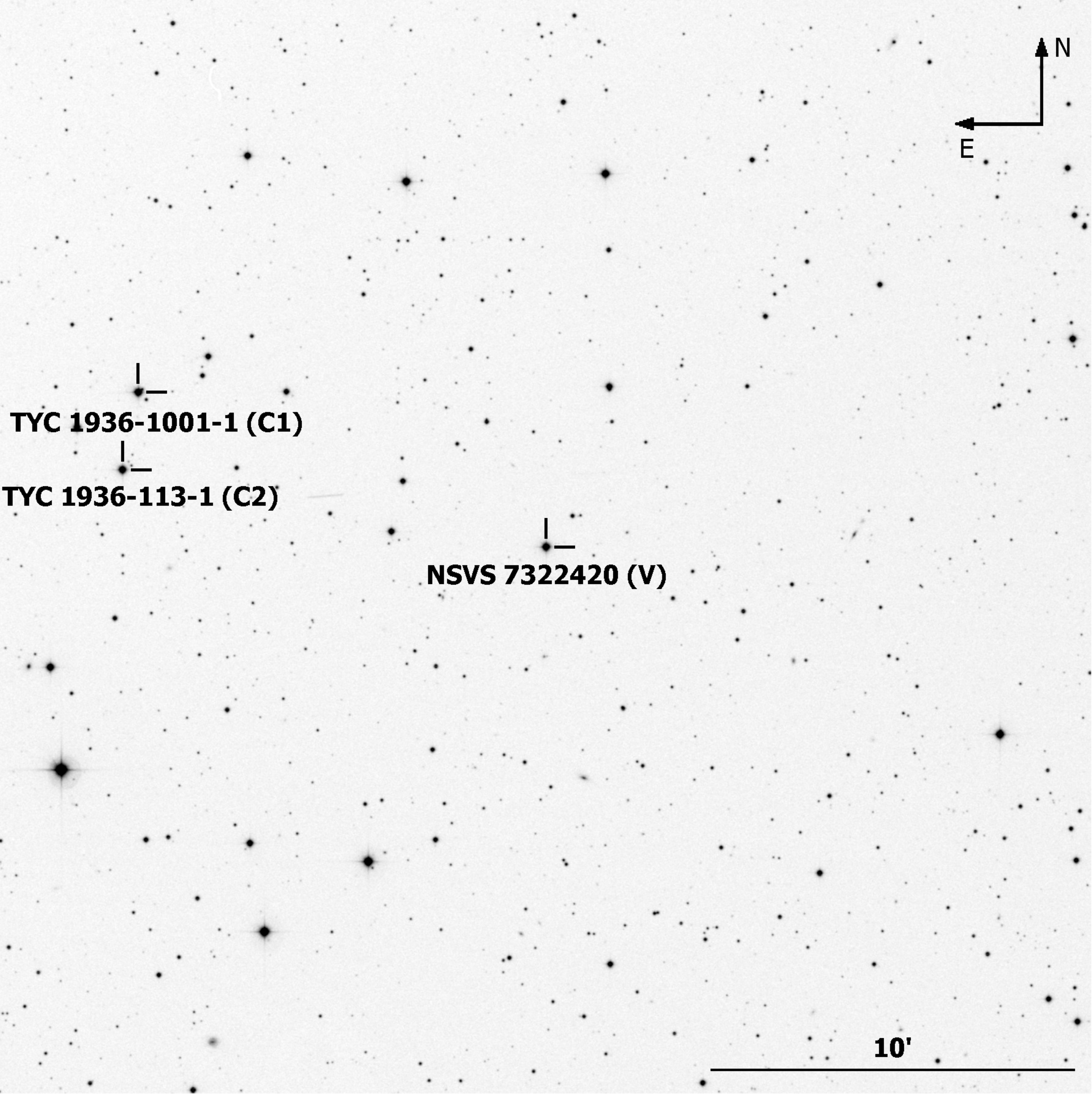}
\includegraphics[width=\columnwidth]{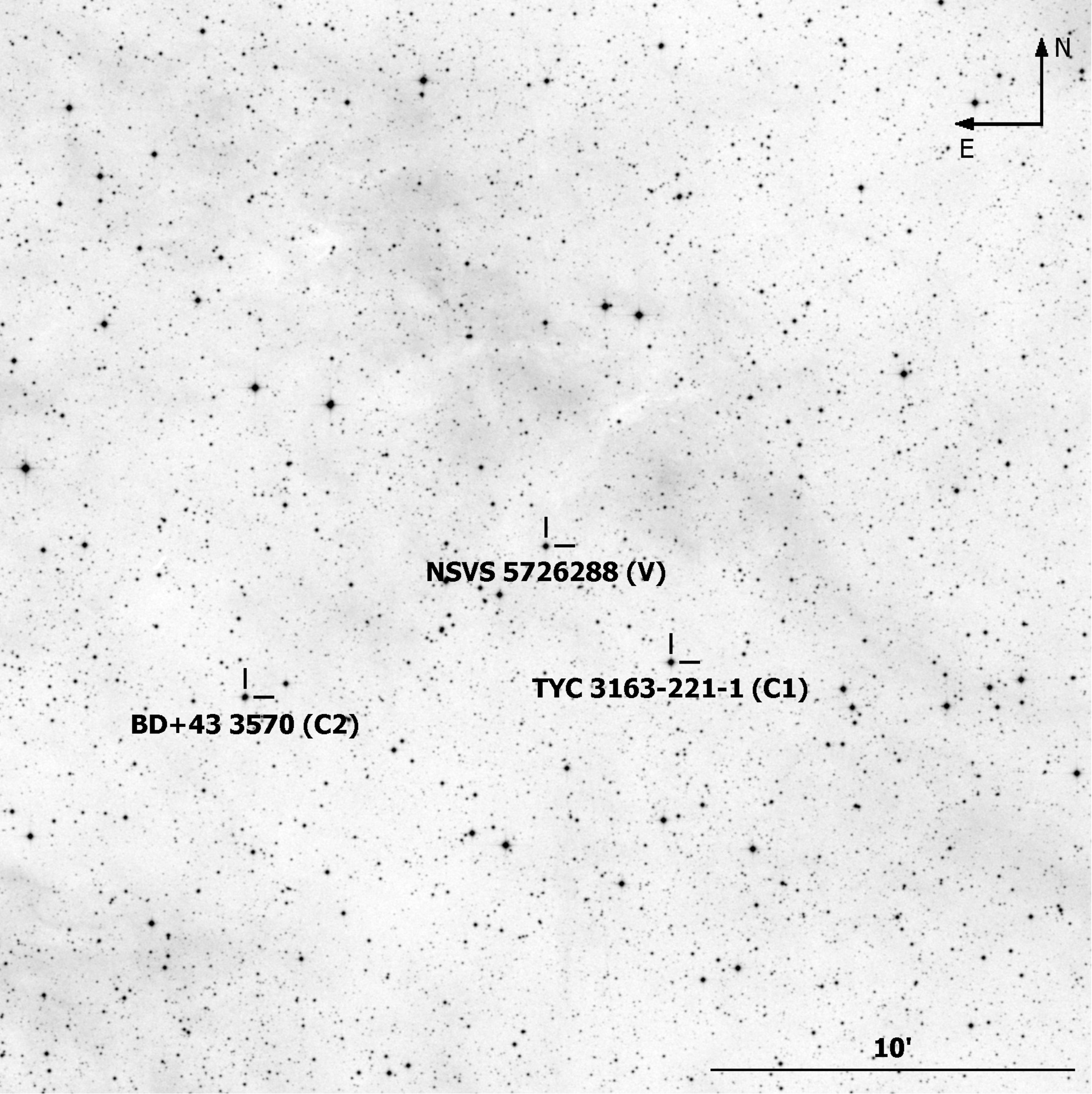}
\caption{Finder charts for NSVS 7322420 (upper panel) and NSVS 5726288 (lower panel) as well as their comparison stars (labeled C1 and C2).}
\label{fig:Finder-charts}
\end{center}
\end{figure}

We chose two eclipsing binary candidates identified by \citet{2008AJ....136.1067H}, who identified 409 candidate Algol- and $\beta$ Lyrae-type eclipsing binary systems in the Northern Sky Variability Survey (NSVS). The NSVS was a survey of the northern sky conducted to record stellar variations of faint objects \citep{2004AJ....127.2436W}. The survey, which was conducted between April 1, 1999 and March 30, 2000 at Los Alamos National Laboratory, imaged 14 million objects north of declination -38$^{\circ}${} with magnitudes between 8 and 15.5. There were typically a few hundred measurements taken for each object, creating a strong base set for searching for variability. Many systems in the NSVS have not had follow up studies performed due to the enormous volume of data, but works such as \citet{2008AJ....136.1067H} have made progress in cataloguing systems based on the NSVS data set itself. The two targets we chose were NSVS 7322420 and NSVS 5726288; details of these two systems and the stars used to analyze them are given in Table \ref{tab:Systems-characteristics}, while finder charts for the systems are shown in Figure \ref{fig:Finder-charts}. The magnitudes given in Table \ref{tab:Systems-characteristics} are taken from the AAVSO Photometric All Sky Survey (APASS) Data Release 9 (\citeauthor{2016yCat.2336....0H}

\newpage

\noindent\citeyear{2016yCat.2336....0H}); the Sloan r' and i' magnitudes given by APASS were converted into Cousins R and I magnitudes using transformation equations found in Table 1 of \citet{2005AJ....130..873J}.

Section \ref{sec:Observations} of this paper describes our acquisition and reduction of the observational data. Section \ref{sec:Analysis} outlines the methods used to analyze our data as well as the results of our analysis. Finally, section \ref{sec:Discussion} contains the discussion of our results as well as details of potential future work.

\section{Observations\label{sec:Observations}}

\begin{table}
\begin{center}
\begin{tabular}{cccc}
\hline 
Target & Observation Date & Telescope & Filters\\
\hline
NSVS 7322420 & 03/22/2013 & BSUO & B, V, R, I\\
 & 03/30/2013 & BSUO & B, V, R, I\\
 & 04/02/2013 & BSUO & B, V, R, I\\
 & 04/03/2013 & BSUO & B, V, R, I\\
 & 04/04/2013 & BSUO & B, V, R, I\\
 & 04/05/2013 & BSUO & B, V, R, I\\
 & 04/14/2013 & BSUO & V, R, I\\
 & 04/16/2013 & KPNO & V, R, I\\
 & 04/21/2013 & BSUO & V, R, I\\
 & 04/22/2013 & BSUO & V, R, I\\
 & 04/30/2013 & BSUO & B, V, R, I\\
 & 05/01/2013 & BSUO & B, V, R, I\\
 & 05/19/2013 & BSUO & B, V, R, I\\
 & 05/27/2013 & KPNO & B, V, R, I\\
 & 11/29/2013 & KPNO & B, V, R, I\\
 & 01/17/2014 & KPNO & B, V, R, I\\
 & 01/28/2014 & KPNO & B, V, R, I\\
 & 01/29/2014 & KPNO & B, V, R, I\\
 & 10/15/2014 & KPNO & B, V, R, I\\
 & 10/27/2014 & KPNO & B, V, R, I\\
\hline 
NSVS 5726288 & 05/15/2013 & BSUO & B, V, R, I\\
 & 05/19/2013 & BSUO & B, V, R, I\\
 & 05/25/2013 & BSUO & B, V, R, I\\
 & 06/03/2013 & BSUO & B, V, R, I\\
 & 06/04/2013 & BSUO & B, V, R, I\\
 & 06/11/2013 & BSUO & B, V, R, I\\
 & 06/14/2013 & BSUO & B, V, R, I\\
 & 06/15/2013 & BSUO & B, V, R, I\\
 & 06/17/2013 & BSUO & B, V, R, I\\
 & 06/19/2013 & BSUO & B, V, R, I\\
 & 06/20/2013 & BSUO & B, V, R, I\\
 & 07/17/2013 & BSUO & B, V, R, I\\
 & 09/22/2014 & BSUO & B, V, R, I\\
 & 09/23/2014 & BSUO & B, V, R, I\\
 & 09/25/2014 & BSUO & B, V, R, I\\
 & 10/15/2014 & KPNO & B, V, R, I\\
 & 05/29/2015 & BSUO & B, V, R\\
 & 11/14/2016 & KPNO & B, V, R, I\\
 & 11/15/2016 & KPNO & B, V, R, I\\
\hline 
\end{tabular}
\caption{A list of dates of observation along with the location and filters used.}
\label{tab:Observation-dates}
\end{center}
\end{table}

\begin{figure}
\begin{center}  
\includegraphics[width=0.9\columnwidth]{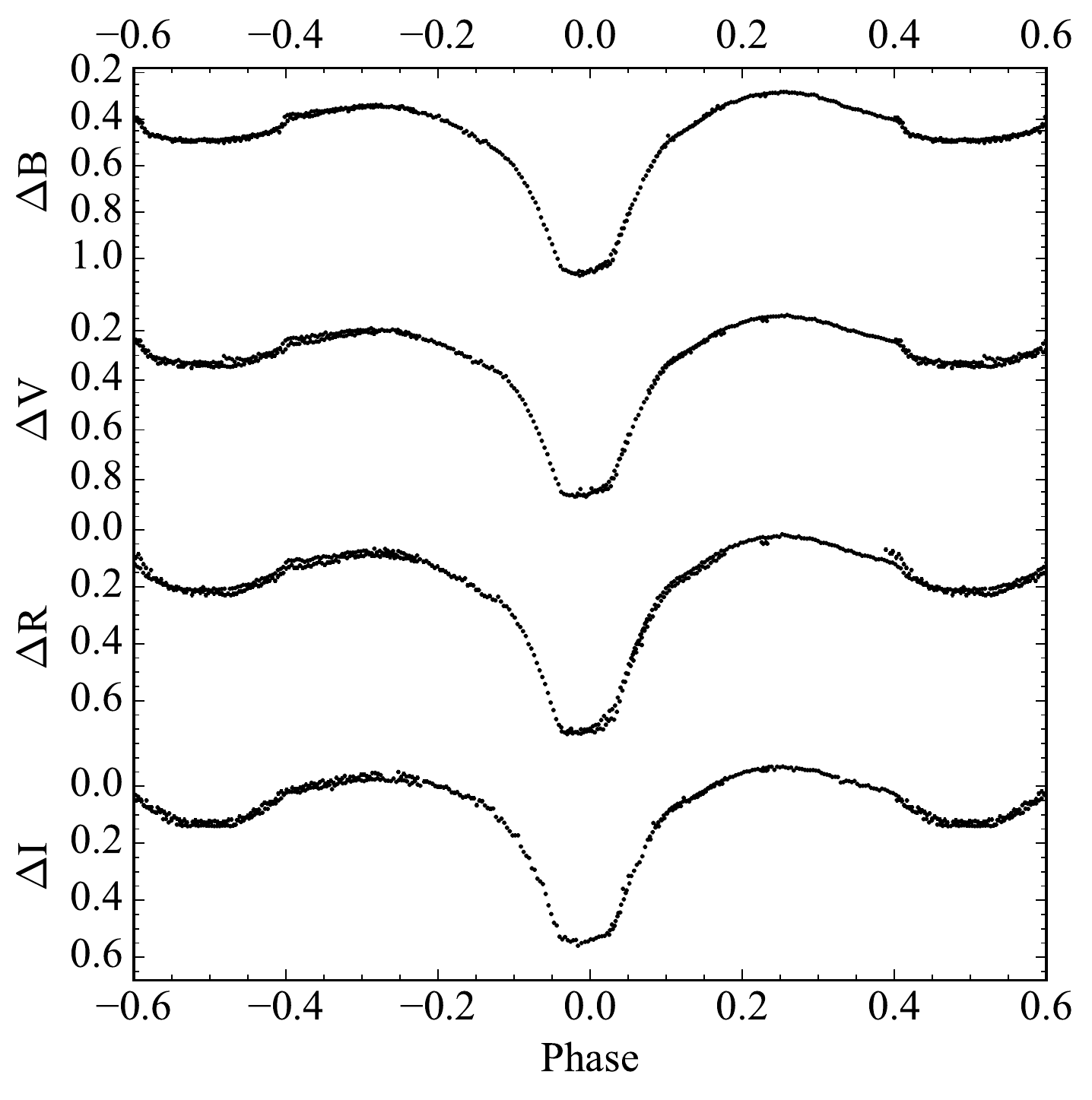}
\includegraphics[width=0.9\columnwidth]{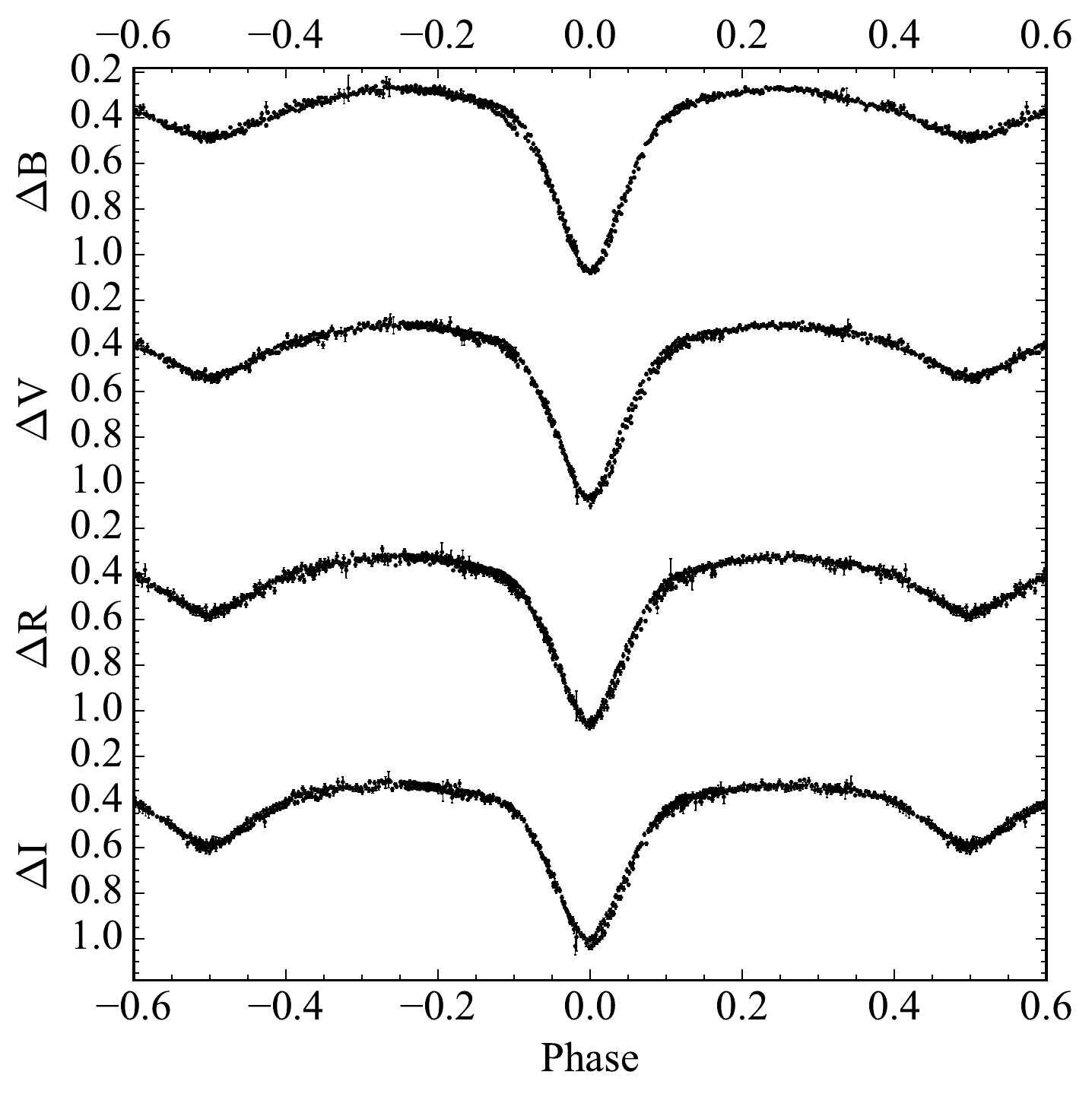}
\caption{Phased light curves for B, V, R, and I-band data on NSVS 7322420 (upper panel) and NSVS 5726288 (lower panel). Error bars are plotted for all data points. The vertical axis shows the differential magnitude in the given band, while the horizontal axis shows the orbital phase of the system. The data for NSVS 7322420 is only the subset of data taken from January 17, 2014 to January 29, 2014.}
\label{fig:Observational-data}
\end{center}
\end{figure}

Our survey of the two systems was conducted between March 2013 and November 2016. We conducted our observations at two separate sites: the Ball State University Observatory (BSUO) in Muncie, Indiana and Kitt Peak National Observatory (KPNO). The BSUO observations were conducted with a STXL-6303E CCD camera cooled to -10 $^{\circ}$C mounted on a 16-inch Meade LX200 while the KPNO observations were conducted with a CCD custom built by Astronomical Research Cameras, Inc (ARC) cooled to -110 $^{\circ}$C mounted on the 0.9-meter SARA-KP telescope \citep{2017PASP..129a5002K}. Observations were conducted in the Bessell B, V, R, and I filters on all nights excepting brief periods in late April 2013 and May 2015; the Bessell filters \citep{1990PASP..102.1181B} closely approximate the Johnson-Cousins photometric system. A list of the nights of  observation is presented in Table \ref{tab:Observation-dates}.

Reduction of the obtained images was performed using the CCDRED package contained in the Image Reduction and Analysis Facility (IRAF, \citealt{1993ASPC...52..173T}). The software AstroImageJ \citep{2017AJ....153...77C}  was then used to perform photometry on the calibrated images. We used the method of differential photometry to obtain the differential magnitude of our system. The comparison star has a similar color index to the variable so that the two will be similarly affected by atmospheric extinction, although the choice of comparison is limited due to the field of view of our instruments (approximately 15 arcminutes). The comparison star is compared to a check star to ensure that the comparison is non-variable. Once the photometry has been performed, the data is converted to a function of orbital phase ($\Phi$) rather than absolute time using Equations \ref{eq:Tar-a-ephemeris-eq} and \ref{eq:Tar-b-ephemeris-eq} for NSVS 7322420 and NSVS 5726288, respectively. The phased light curves that were produced and used in our analysis are shown in Figure \ref{fig:Observational-data}. The unphased standardized magnitude data was uploaded to the AAVSO International Database \citep{stella_kafka_2015} where it is available for download.

\section{Analysis\label{sec:Analysis}}

\subsection{Methodology}

\subsubsection{Ephemeris, period determination, and O$-$C calculation}

\begin{table*}
\begin{center}
\begin{tabular}{ccccc}
\hline 
Target & Epoch & \multicolumn{2}{c}{Time of Minimum (HJD)} & O$-$C (days)\\
       &       & Observed & Calculated &  \\
\hline
NSVS 7322420 & 0 & $2456413.632618 \pm{} 0.000201$ & --- & $0 \pm 0.000201$\\
 & 454 & $2456625.861477 \pm 0.000027$ & $2456625.862636 \pm 0.000201$ & $-0.001159 \pm 0.000203$\\
 & 559 & $2456674.944067 \pm 0.000100$ & $2456674.946671 \pm 0.000201$ & $-0.002604 \pm 0.000225$\\
\hline 
NSVS 5726288 & 0 & $2456457.738889 \pm 0.000178$ & --- & $0 \pm 0.000178$\\
 & 543 & $2456922.677753 \pm 0.000231$ & $2456922.685354 \pm 0.000178$ & $-0.007601 \pm 0.000292$\\
 & 570 & $2456945.796265 \pm 0.000127$ & $2456945.804239 \pm 0.000178$ & $-0.007974 \pm 0.000219$\\
\hline 
\end{tabular}
\caption{O$-$C and observed and calculated times of primary minimum for NSVS 7322420 and NSVS 5726288.}
\label{tab:Ephimeris-table}
\end{center}
\end{table*}

\begin{figure}
\begin{center}  
\includegraphics[width=0.85\columnwidth]{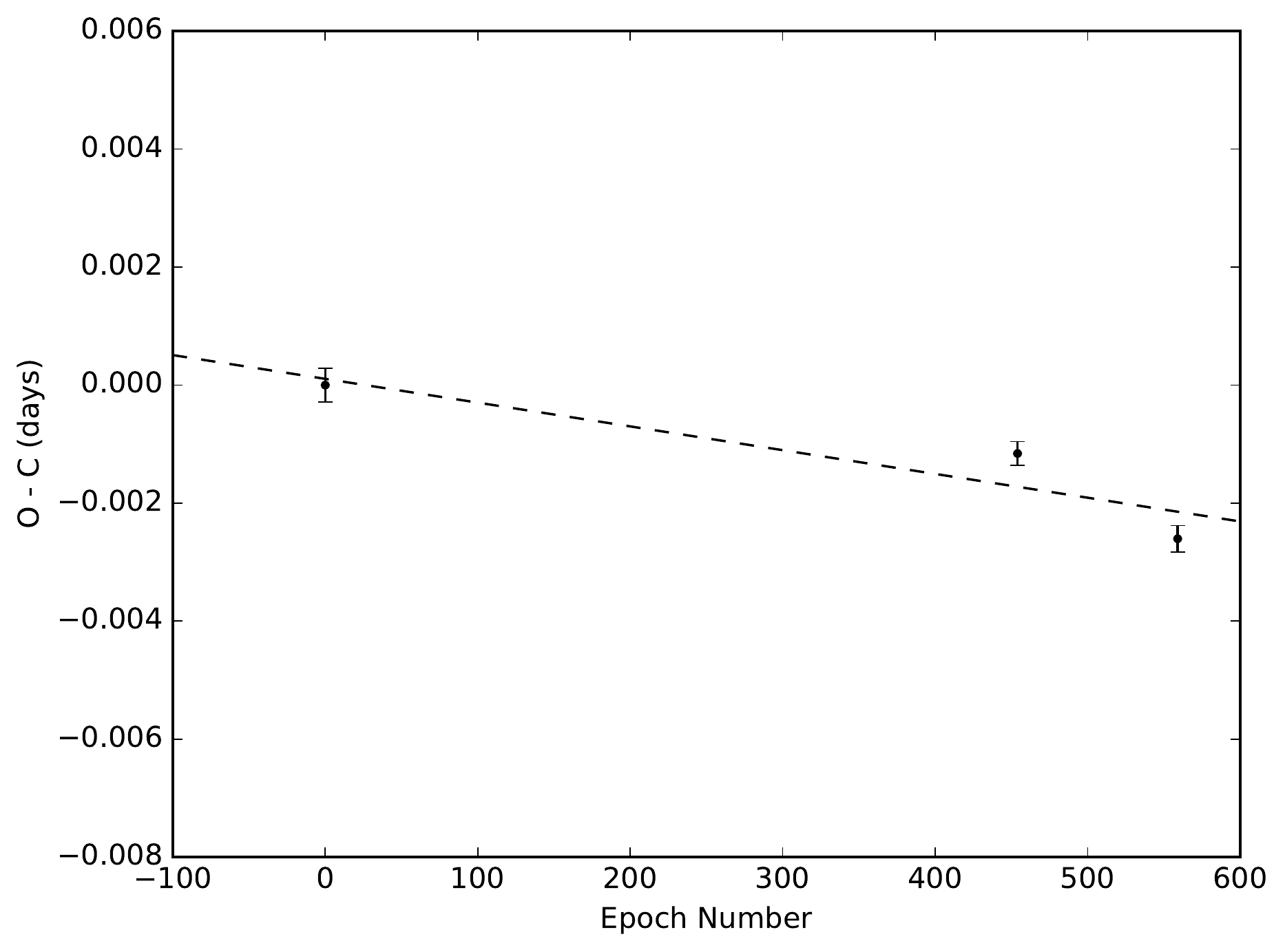}
\includegraphics[width=0.85\columnwidth]{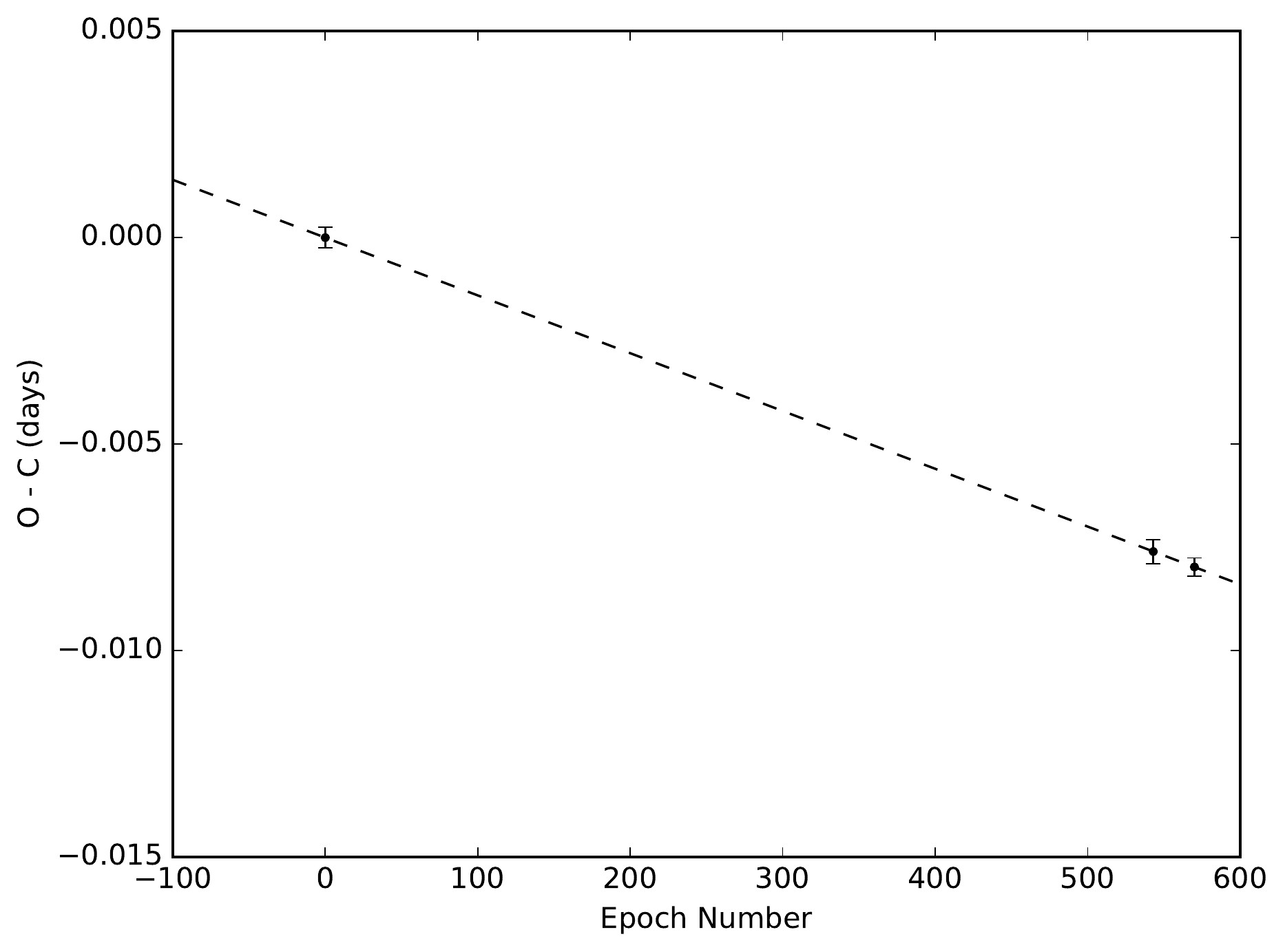}
\caption{Observed minus calculated (O $-$ C) times of minimum light for NSVS 7322420 (upper panel) and NSVS 5726288 (lower panel) plotted against epoch number. The dashed lines indicate the best linear fit to the data.}
\label{fig:O-C-plot}
\end{center}
\end{figure}

\begin{figure}
\begin{center}
\includegraphics[width=\columnwidth]{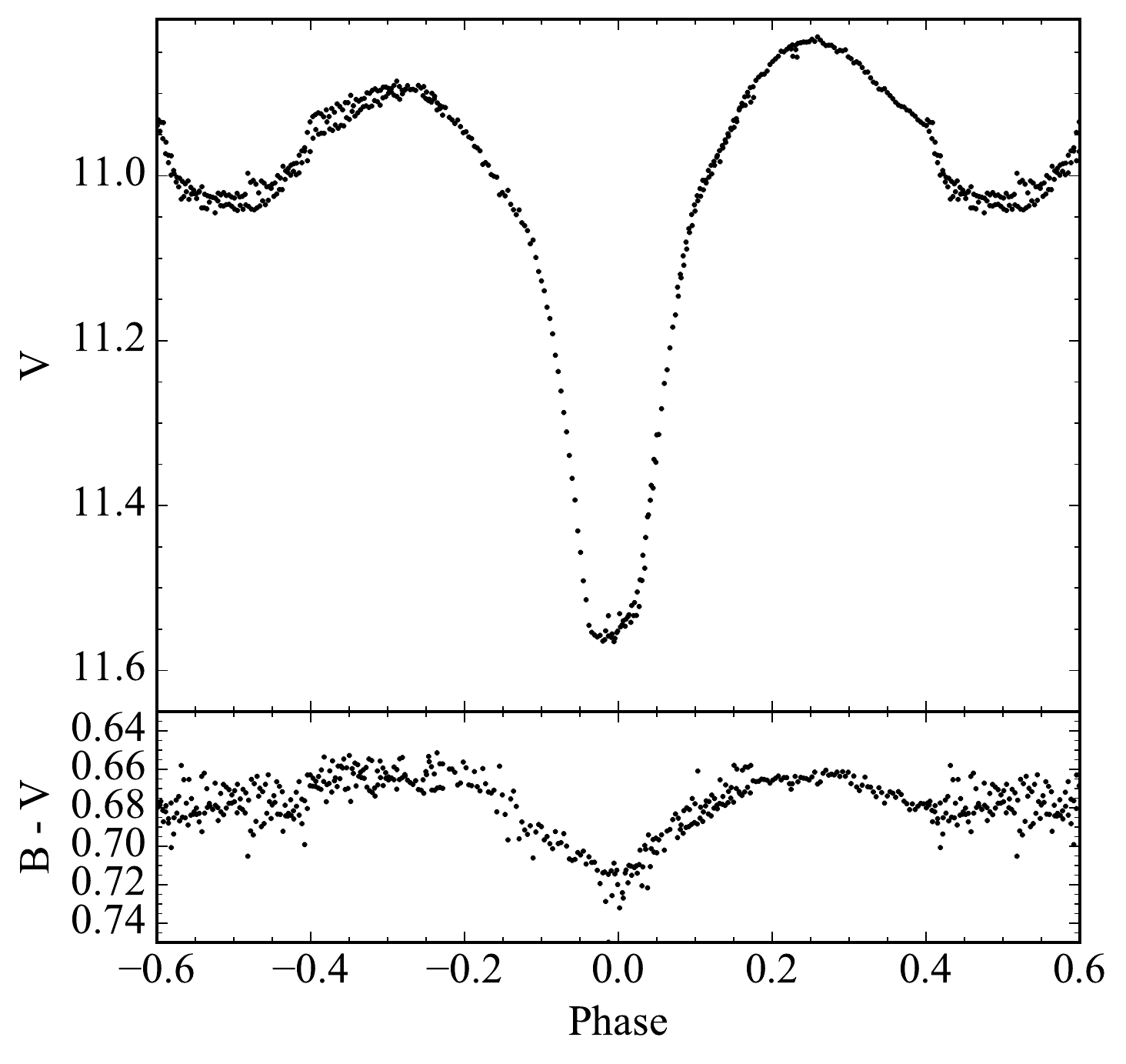}
\caption{(B$-$V) color curve for NSVS 7322420. The apparent V-band magnitude is plotted in the top panel, while the (B$-$V) color curve is plotted in the lower panel. Error bars are not plotted for clarity.}
\label{fig:Tar-a-b-v}
\end{center}
\end{figure}

\begin{figure}[h!]
\begin{center}
\includegraphics[width=\columnwidth, trim= 150 300 90 300]{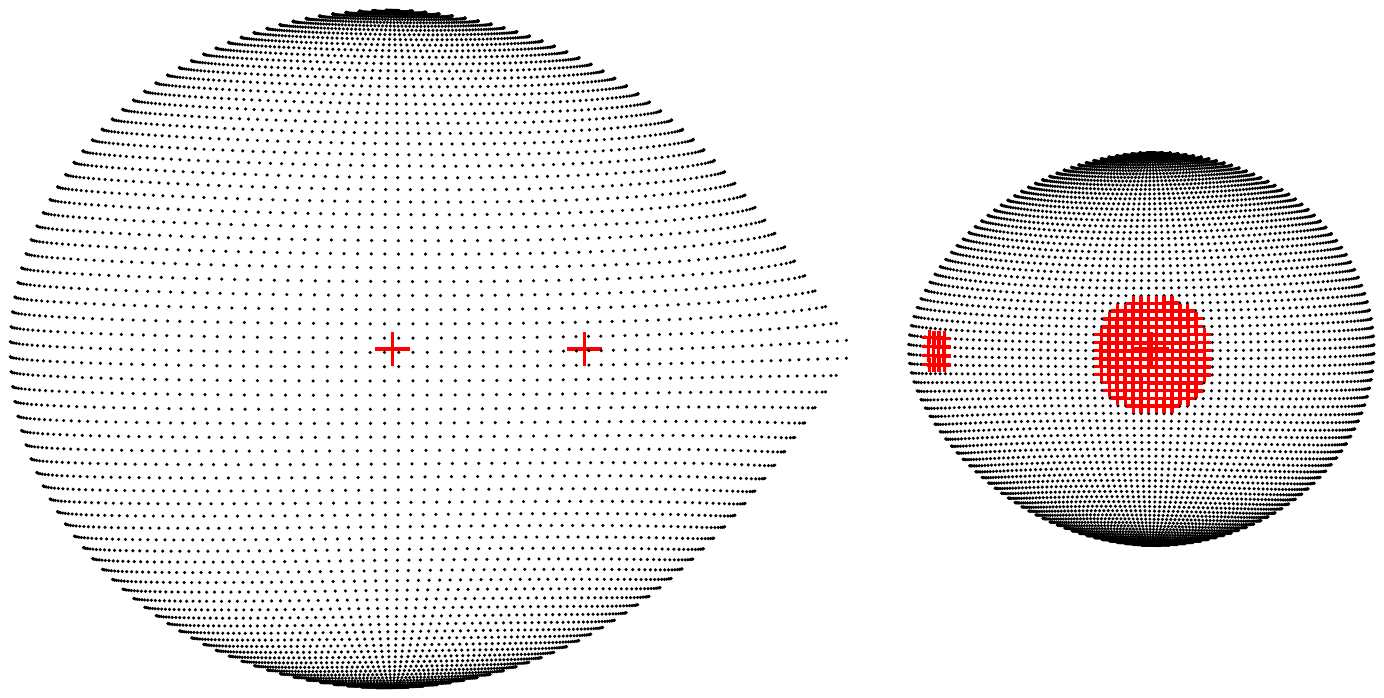}
\caption{Three-dimensional visualization of the model for NSVS 7322420. The phase of the system in this figure is $0.25$, with the primary component on the left and the secondary component on the right; the primary is moving toward the observer. The red markers on the left and right represent the center of each star while the red marker in the center represents the center of mass of the binary system. Two of the three hot spots are clearly visible on the secondary component.}
\label{fig:Tar-a-3d-model}
\end{center}
\end{figure}

The time of minimum for each primary eclipse was calculated using the method described by \citet{1956BAN....12..327K}. This process was applied to each eclipse in each filter, and the obtained values for each individual eclipse were averaged together to give the reported time of minimum. Period determination was done using the program Peranso \citep{2016AN....337..239P} rather than by examination of the times of minimum light. Peranso allows the user to input a set of observations and use one of a variety of methods to calculate the period. We chose to use the analysis of variance (ANOVA) method described by \citet{1996ApJ...460L.107S}. This method, which uses periodic orthogonal polynomials to fit the observed data, excels at detecting and discarding aliases of the true period.

The observed times of minimum light were then compared to a calculated time of minimum. This calculated value for the $\rm n^{th}$ epoch is found using the equation:

\begin{equation}
\rm T_{cal,n}=T_{obs,0}+E_nP
\label{eq:o-c}
\end{equation}
where $\rm T_{obs,0}$ is an observed reference time of minimum, $\rm E_n$ is the number of periods elapsed since $\rm T_{obs,0}$, and $\rm P$ is the orbital period. The calculated value was then obtained by subtracting it from the corresponding observed value to obtain O$-$C. The error in O$-$C is given by the equation:

\begin{equation}
\rm \sigma_{{\rm O-C},n}=\sqrt{\sigma_{T_{{\rm obs},n}}^{2}+\sigma_{T_{{\rm obs},0}}^{2}}
\end{equation}
where $\rm \sigma$ denotes the error of the subscripted parameter.

\subsubsection{Magnitude calibration}

\begin{table*}
\begin{center}
\begin{tabular}{ccc}
\hline 
Parameter & NSVS 7322420 & NSVS 5726288\\
\hline
Orbital Period (d) & 0.467467 $\pm$ 0.000015 & 0.856255 $\pm$ 0.000007\\
T$_{\rm eff}$ of Primary Component (K, fixed) & 5,700 & 7,300\\
T$_{\rm eff}$ of Secondary Component (K) & 3,361 $\pm$ 29 & 5,632 $\pm$ 37\\
Orbital Inclination ($^{\circ}$) & 91.66 $\pm$ 0.04 & 79.60 $\pm$ 0.02\\
Surface Potential of Primary Component & --- & 3.1608 $\pm$ 0.0035\\
Surface Potential of Secondary Component & 2.6085 $\pm$ 0.0088 & 3.1791 $\pm$ 0.0039\\
Mass Ratio & 0.3388 $\pm$ 0.0025 & 0.5789 $\pm$ 0.0014\\
Magnitude Difference ($M_{2}-M_{1}$) & 3.4974 & 1.7431\\
Spot 1 Temperature (K) & 6,227 $\pm$ 89 & --- \\
Spot 1 Latitude ($^{\circ}$, fixed) & 0 & --- \\
Spot 1 Longitude ($^{\circ}$, fixed) & 90 & --- \\
Spot 1 Radius ($^{\circ}$, fixed) & 16 & --- \\
Spot 2 Temperature (K, fixed) & 6,722 $\pm$ 58 & --- \\
Spot 2 Latitude ($^{\circ}$, fixed) & 0 & --- \\
Spot 2 Longitude ($^{\circ}$, fixed) & 30 & --- \\
Spot 2 Radius ($^{\circ}$, fixed) & 5 & --- \\
Spot 3 Temperature (K, fixed) & 6,722 $\pm$ 58 & --- \\
Spot 3 Latitude ($^{\circ}$, fixed) & 0 & --- \\
Spot 3 Longitude ($^{\circ}$, fixed) & 340 & --- \\
Spot 3 Radius ($^{\circ}$, fixed) & 5 & --- \\
\hline 
\end{tabular}
\caption{Results of the modeling process and analysis for NSVS 7322420 and NSVS 5726288. The errors are the formal errors given by PHOEBE.}
\label{tab:Model-results}
\end{center}
\end{table*}

The output file generated by AstroImageJ gives the brightness of the system in relative flux, from which the differential magnitude can be calculated. Estimating the temperature of the stellar components, however, requires the apparent magnitude of the system. The apparent magnitude was obtained by adding the calibrated apparent magnitude of the comparison star given by \citet{2016yCat.2336....0H} to the differential magnitude obtained from AstroImageJ. We performed an analysis to determine the effect the atmosphere had on our determination of the magnitude; we found that the correction factor was negligible and thus our determined magnitude very closely approximates the true apparent magnitude. The error in magnitude was then calculated by adding in quadrature the error in the apparent magnitude of the comparison star (also provided by \citeauthor{2016yCat.2336....0H}) and the error determined from AstroImageJ. We performed the magnitude calibration and error determination on all data points.

\subsubsection{Temperature estimation}

A rough estimation for the temperature of the primary component was made by calculating the (B$-$V) color index of the system at phase $0.5$. Since the observations in the B and V filters did not occur simultaneously, the observational values from the B filter were linearly interpolated to coincide with the values for the V filter. The difference between the interpolated B data and observed V data were then calculated and displayed as a function of phase. These raw values were corrected for interstellar reddening based on the work by \citet{2011ApJ...737..103S}, which provides E(B$-$V) values for given coordinates based on the assumption that we are observing objects beyond our galaxy. While this assumption does not hold for our stars, this provides a first approximation correction.

Once the corrected (B$-$V) color curve was obtained, the temperature of the primary component was estimated based upon the (B$-$V) color index at phase $0.5$. At this phase, the primary component mostly or completely obscures the secondary and is therefore the dominant contributor to the observed flux. We used the work presented in \citet{1996ApJ...469..355F} to convert the (B$-$V) color index at phase 0.5 to temperature under the assumption that the stars are on the main sequence. The secondary temperature was determined through modeling.

\subsubsection{Light curve analysis}

We used the program PHOEBE (PHysics Of Eclipsing BinariEs) v0.31a \citep{2005ApJ...628..426P} to analyze the light curves of the systems and produce a best-fit model. PHOEBE is a graphical user interface built on the Wilson-Devinney (WD) program introduced by \citet{1971ApJ...166..605W}. This model includes many parameters that affect the synthetic light curve produced by the program. These parameters are altered using an iterative least-squares analysis known as differential correction to produce the best fit to the observed data.

We input the phased apparent magnitude curves from all four filters into PHOEBE. Data from all filters were fit simultaneously to better constrain the system parameters. The primary temperature was held fixed at the previously estimated value throughout the modeling process while the secondary temperature was allowed to vary. The fact that all of our target stars had temperatures suggesting convective envelopes allowed us to set the gravity brightening and surface albedos to their theoretical values of 0.32 \citep{1967ZA.....65...89L} and 0.5 \citep{1969AcA....19..245R}, respectively. We performed a mass ratio search (as described in \citealt{2007ApJ...671..811Q}) to determine the best fit mass ratio.

Following this, we allowed other parameters of the system to vary. These parameters include the Kopal surface potential ($\Omega$), the mass ratio of the system ($q = \rm m_2 / m_1$), the orbital inclination ($i$) of the system, and a phase shift (which  was negligible for both systems). Throughout the process, we continued to normalize the luminosities of the component stars and interpolate limb darkening coefficients based on the work by \citet{1993AJ....106.2096V}. The limb darkening coefficients were interpolated using a logarithmic law as the temperature for all stars is less than 9,000 K.

\subsection{NSVS 7322420}

Our period analysis for NSVS 7322420 yielded a period of $0.467467 \pm0.000015$ days, which is quite close to the value of $0.46740$ days published by \citet{2008AJ....136.1067H}. We determined the average time of minimum on three primary eclipses, and these and the O$-$C values are listed in Table \ref{tab:Ephimeris-table} while a diagram showing the best linear fit to these values is displayed in Figure \ref{fig:O-C-plot}. We used the first time of minimum as the reference minimum, resulting in a linear ephemeris of:

\begin{equation}
\rm T_{{min}} (HJD) = 2456413.6326(2) + 0\overset{d}{.}467467(15)E
\label{eq:Tar-a-ephemeris-eq}
\end{equation}
The (B$-$V) color curve we obtained for this system is shown in Figure \ref{fig:Tar-a-b-v}, from which we estimate a (B$-$V) color index of $0.68 \pm 0.07$ during the secondary eclipse (the large error is due to the uncertainty in the apparent magnitudes of the comparison star). \citet{2011ApJ...737..103S} estimate a E(B$-$V) of $0.03$ at the coordinates of this system, giving a (B$-$V) color index of $0.65 \pm 0.07$ during the secondary eclipse. This corresponds to a temperature for the primary component of $5,700 \pm 230$ K and indicates a spectral type of G$3 \pm 3$ \citep{1970A&A.....4..234F}.

\begin{figure}
\begin{center}
\includegraphics[width=\columnwidth]{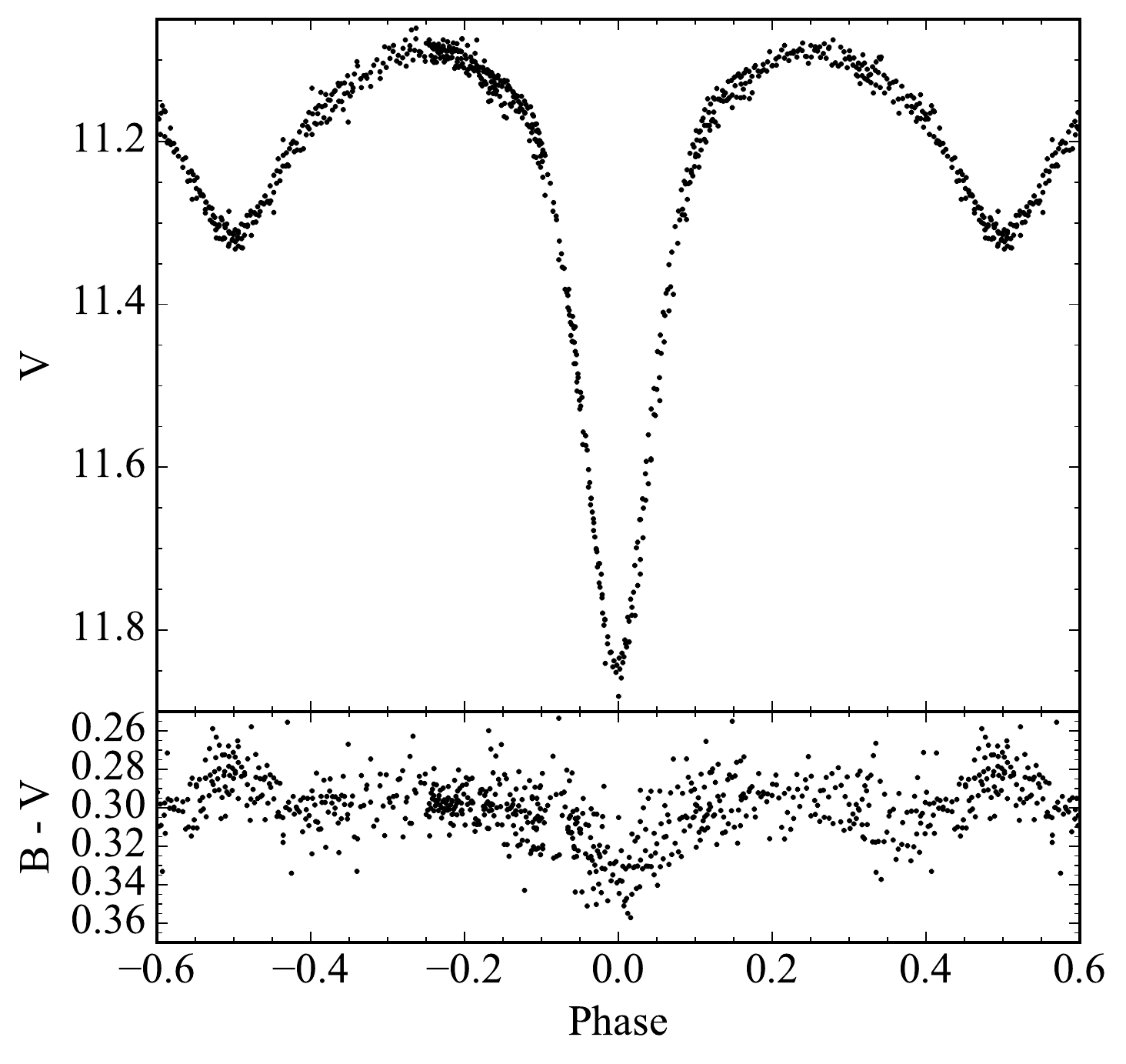}
\caption{(B$-$V) color curve for NSVS 5726288. The apparent V-band magnitude is plotted in the top panel, while the (B$-$V) color curve is plotted in the lower panel. Error bars are not plotted for clarity.}
\label{fig:Tar-b-b-v}
\end{center}
\end{figure}

\begin{figure}
\begin{center}
\includegraphics[width=\columnwidth, trim= 130 310 100 310]{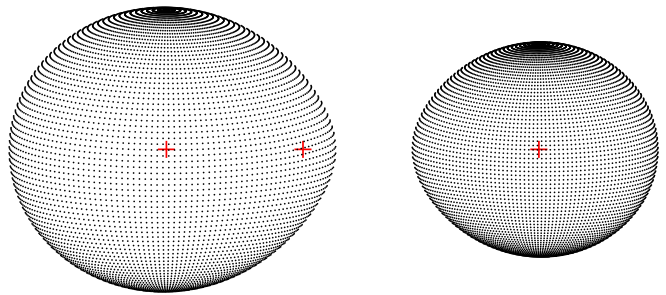}
\caption{Three-dimensional visualization of the model for NSVS 5726288. The phase of the system in this figure is $0.25$, with the primary component on the left and the secondary component on the right; the primary is moving toward the observer. The red markers on the left and right represent the center of each star while the red marker in the center represents the center of mass of the binary system.}
\label{fig:Tar-b-3d-model}
\end{center}
\end{figure}

The system proved difficult to model, due at least in part to the presence of maxima of unequal height: the maximum following primary eclipse is 0.062 magnitudes brighter than the maximum following secondary eclipse in the B band. This is known as the O'Connell effect (\citealt{1951PRCO....2...85O}, \citealt{1968AJ.....73..708M}) and may be explained by a hot or cool spot on one of the stars or from gas streams creating one or more hot spots \citep{2009SASS...28..107W}. The model we present matches the observed light curve marginally well, but it is highly artificial as it requires three hot spots in order to adequately explain the observed features in the light curve. While the hot spot located at 90$^{\circ}$ longitude can be interpreted as the impact of a matter stream onto the stellar surface \citep{2009SASS...28..107W}, the two spots located close to 0$^{\circ}$ longitude have no clear physical interpretation. Spots introduce a considerable amount of degeneracy into the solution, and only the temperature factor of the first spot could be allowed to vary without the program diverging. As a result of these factors, this model should not be interpreted as a completely accurate physical description of the system, and while the general system characteristics are most likely accurate, the specific cause of the light curve irregularities is likely not accurately explained by this model.

\begin{figure}
\begin{center}
\includegraphics[width=0.65\columnwidth]{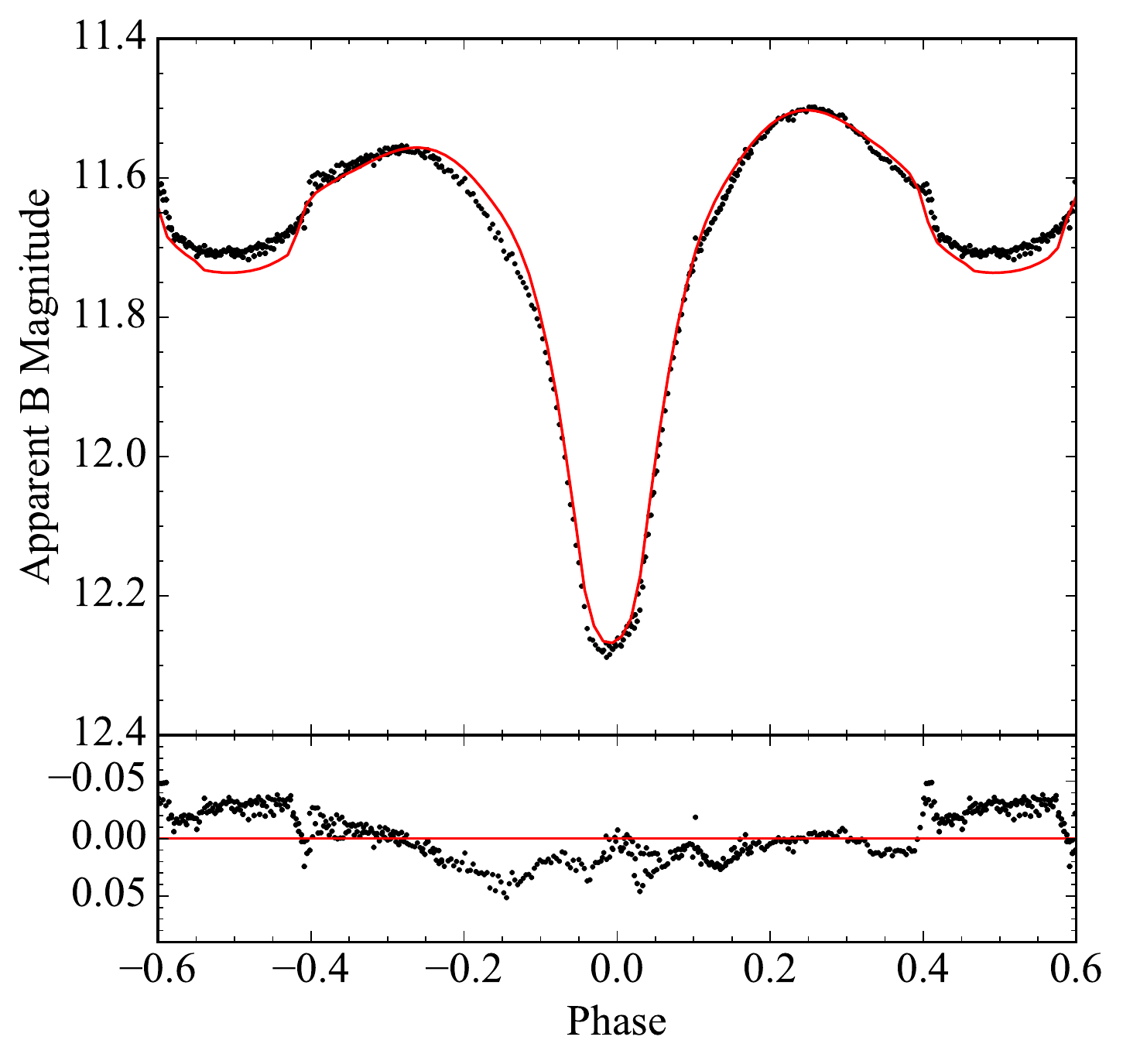}
\includegraphics[width=0.65\columnwidth]{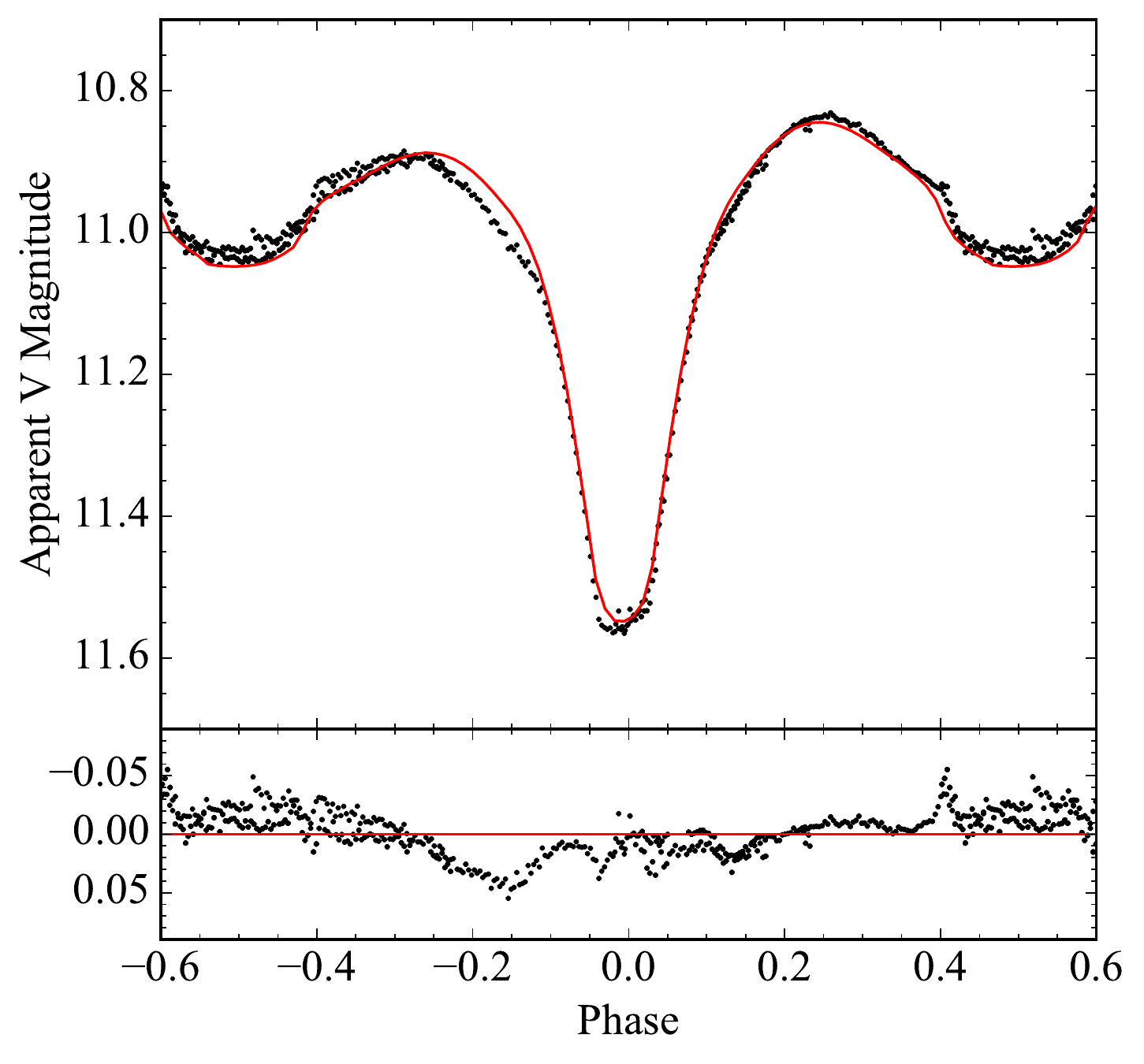}
\includegraphics[width=0.65\columnwidth]{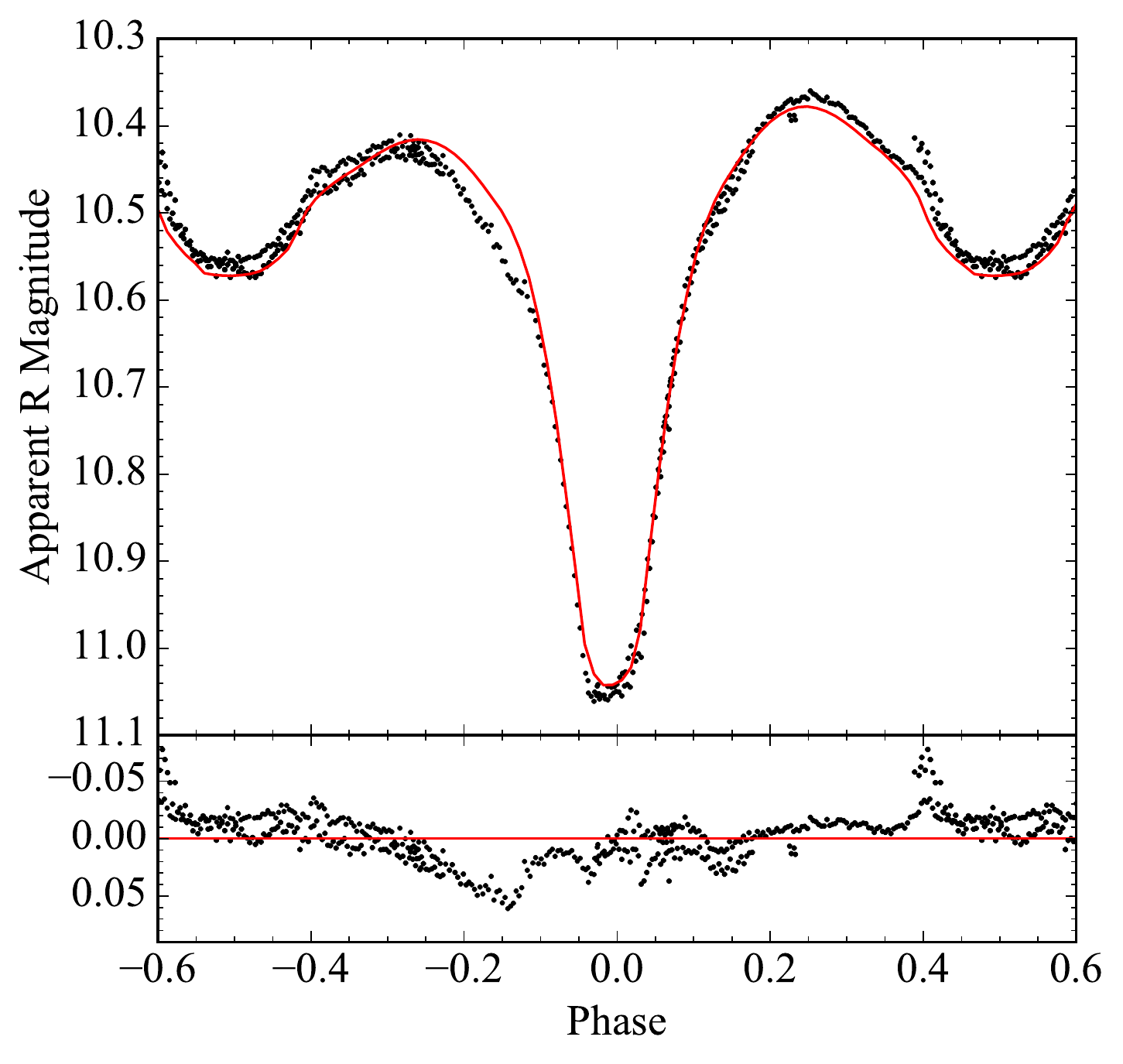}
\includegraphics[width=0.65\columnwidth]{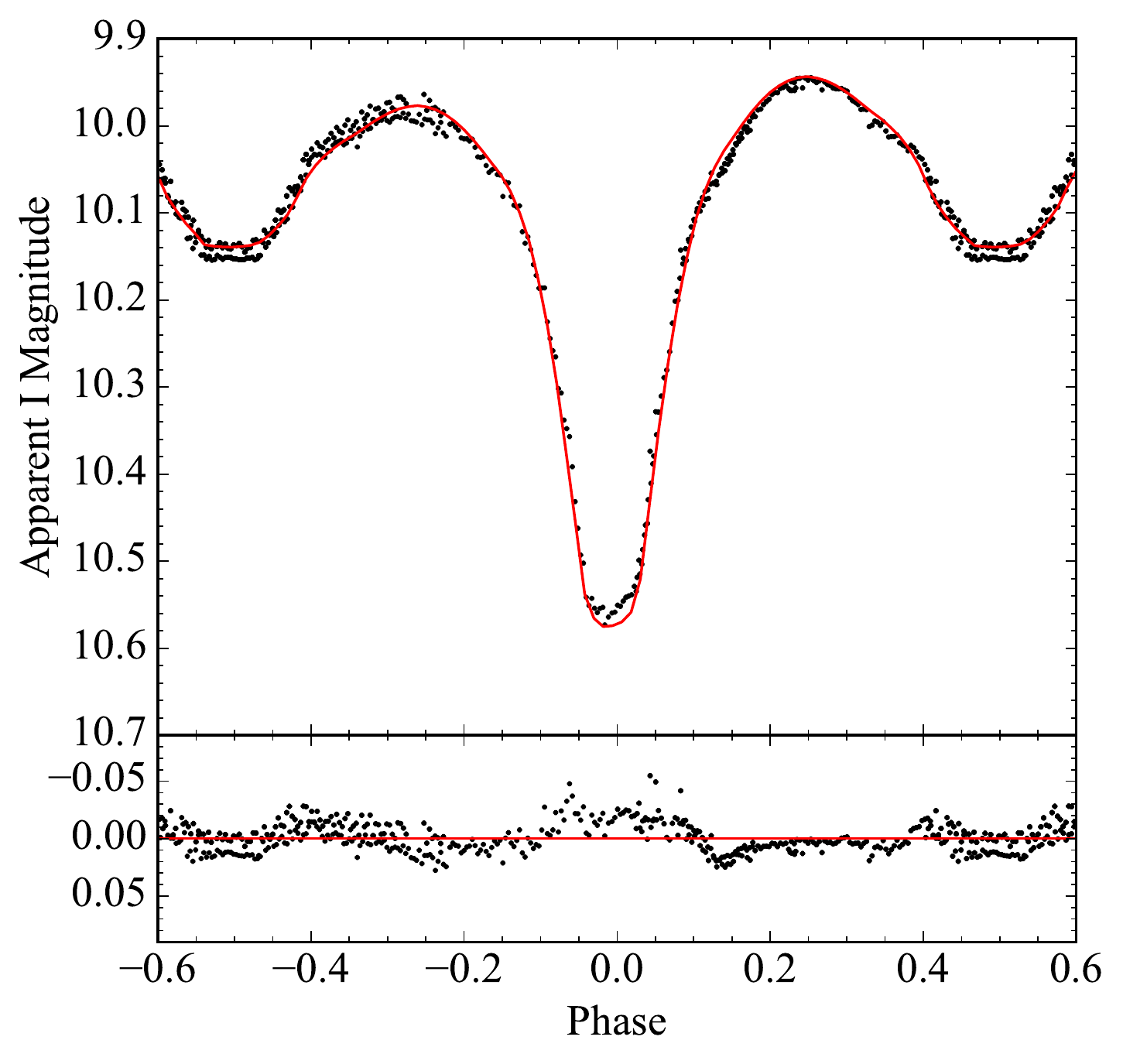}
\caption{Observed B (top panel), V (second panel), R (third panel), and I (fourth panel) data for NSVS 7322420 plotted against synthetic light curves. The top panel of each figure plots the apparent magnitude of the observed and synthetic light curves against the phase of the system. The residuals for the model are shown in the bottom panel of each figure.}
\label{fig:Tar-a-BVRI-models}
\end{center}
\end{figure}

The model indicates that the system is a primary filling semi-detached system with a secondary component of spectral type M. Details of this model is given in the middle column of Table \ref{tab:Model-results} while a three-dimensional visualization (produced by \textit{Binary Maker 3,} \citealt{2002AAS...201.7502B}) of the model is given in Figure \ref{fig:Tar-a-3d-model}. A comparison of the observed data and synthetic light curves is presented in Figure \ref{fig:Tar-a-BVRI-models}.

\subsection{NSVS 5726288}

Our period analysis for NSVS 5726288 yielded a period of $0.856255 \pm0.000007$ days, which differs quite significantly from the value of $0.59935$ days published by \citet{2008AJ....136.1067H}. This 0.59935 day period is almost exactly seven-tenths of our calculated period, indicating that it is likely an alias caused by poor temporal coverage. We determined the average time of minimum on three primary eclipses, and these and the O$-$C values are listed in Table \ref{tab:Ephimeris-table}  while a diagram showing the best linear fit to these values is displayed in Figure \ref{fig:O-C-plot}. We used the first time of minimum as the reference minimum, resulting in a linear ephemeris of:

\begin{equation}
\rm T_{{min}} (HJD) = 2456457.738889(178) + 0\overset{d}{.}856255(7)E
\label{eq:Tar-b-ephemeris-eq}
\end{equation}
The (B$-$V) color curve we obtained for this system is shown in Figure \ref{fig:Tar-b-b-v}, from which we estimate a (B$-$V) color index of $0.28 \pm 0.04$ during the secondary eclipse. \citet{2011ApJ...737..103S} estimate a E(B$-$V) of $1.55$ at the coordinates of this system, giving a (B$-$V) color index of $-1.27 \pm 0.04$ during the secondary eclipse. This gives a non-physical value for the effective temperature, preventing us from using this method to determine an interstellar extinction correction for this system. The reason for such a large correction is that the system lies in the plane of the Milky Way and is therefore subject to significant extinction. With no better way to determine the reddening of the system, we chose to use the raw color index $(0.28 \pm 0.04)$ as a basis for temperature estimation. This corresponds to a temperature of $7,300 \pm 210$ K and indicates a spectral type of A$8 \pm 1$ \citep{1970A&A.....4..234F}.

\begin{figure}
\begin{center}
\includegraphics[width=0.65\columnwidth]{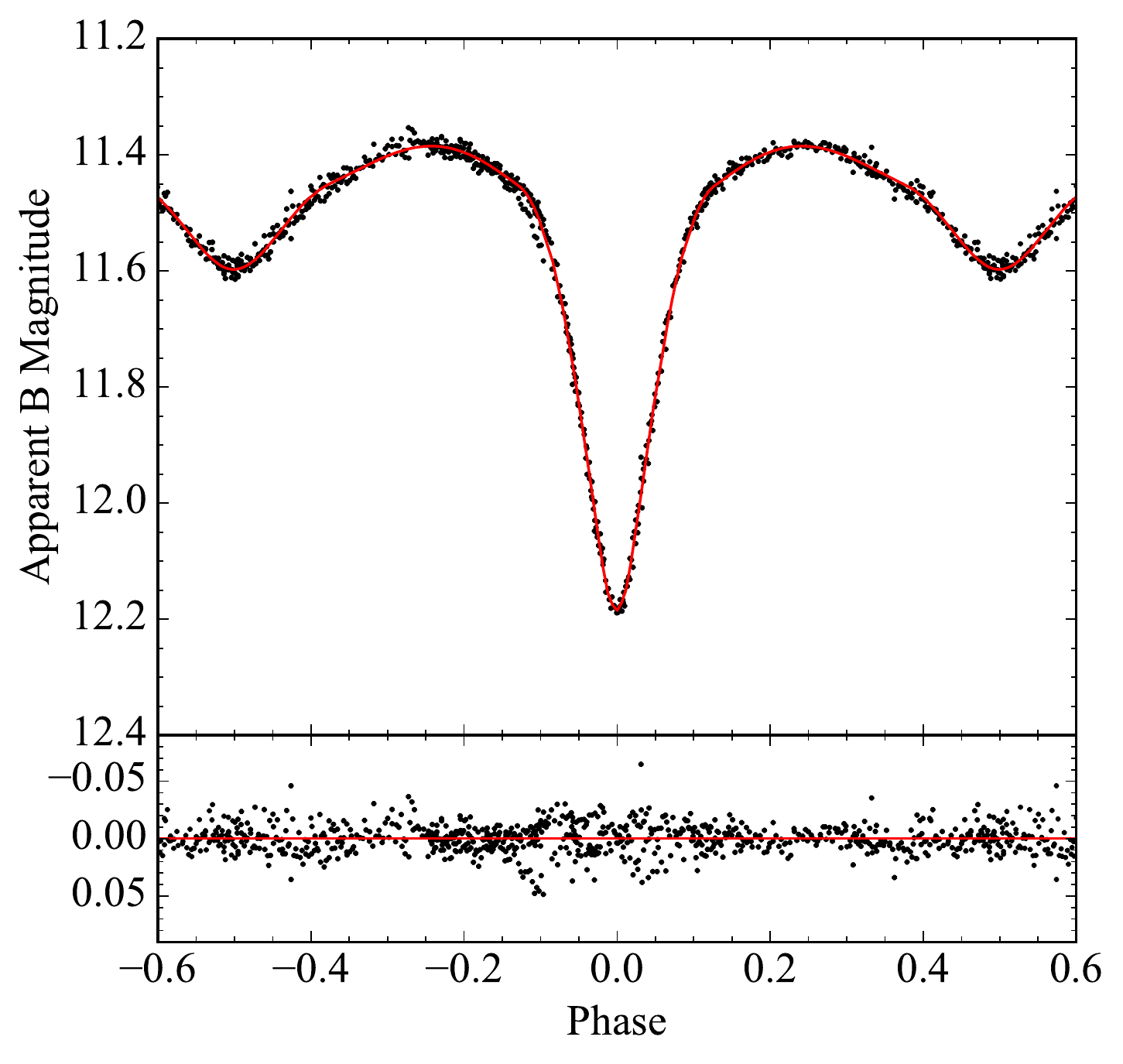}
\includegraphics[width=0.65\columnwidth]{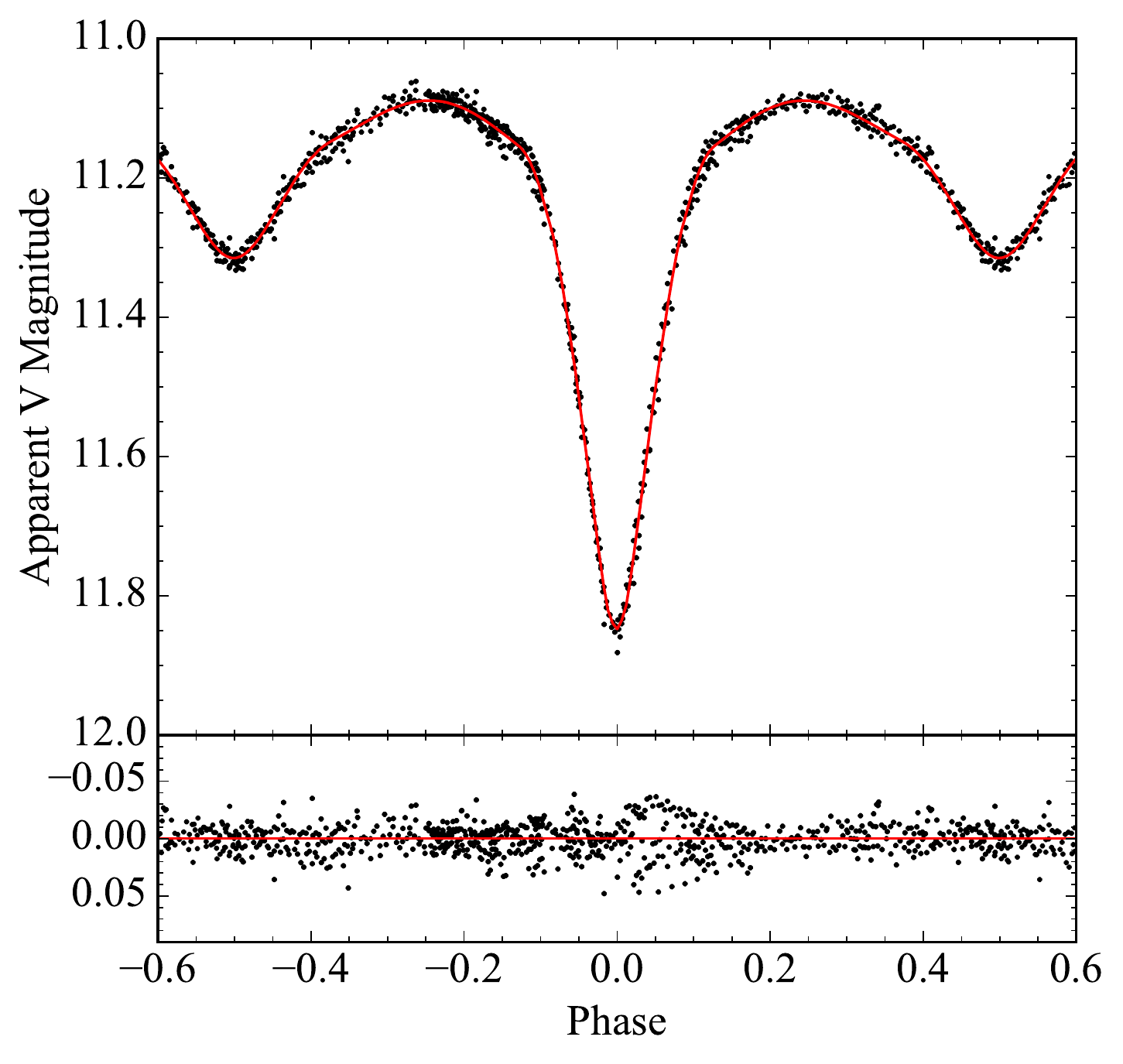}
\includegraphics[width=0.65\columnwidth]{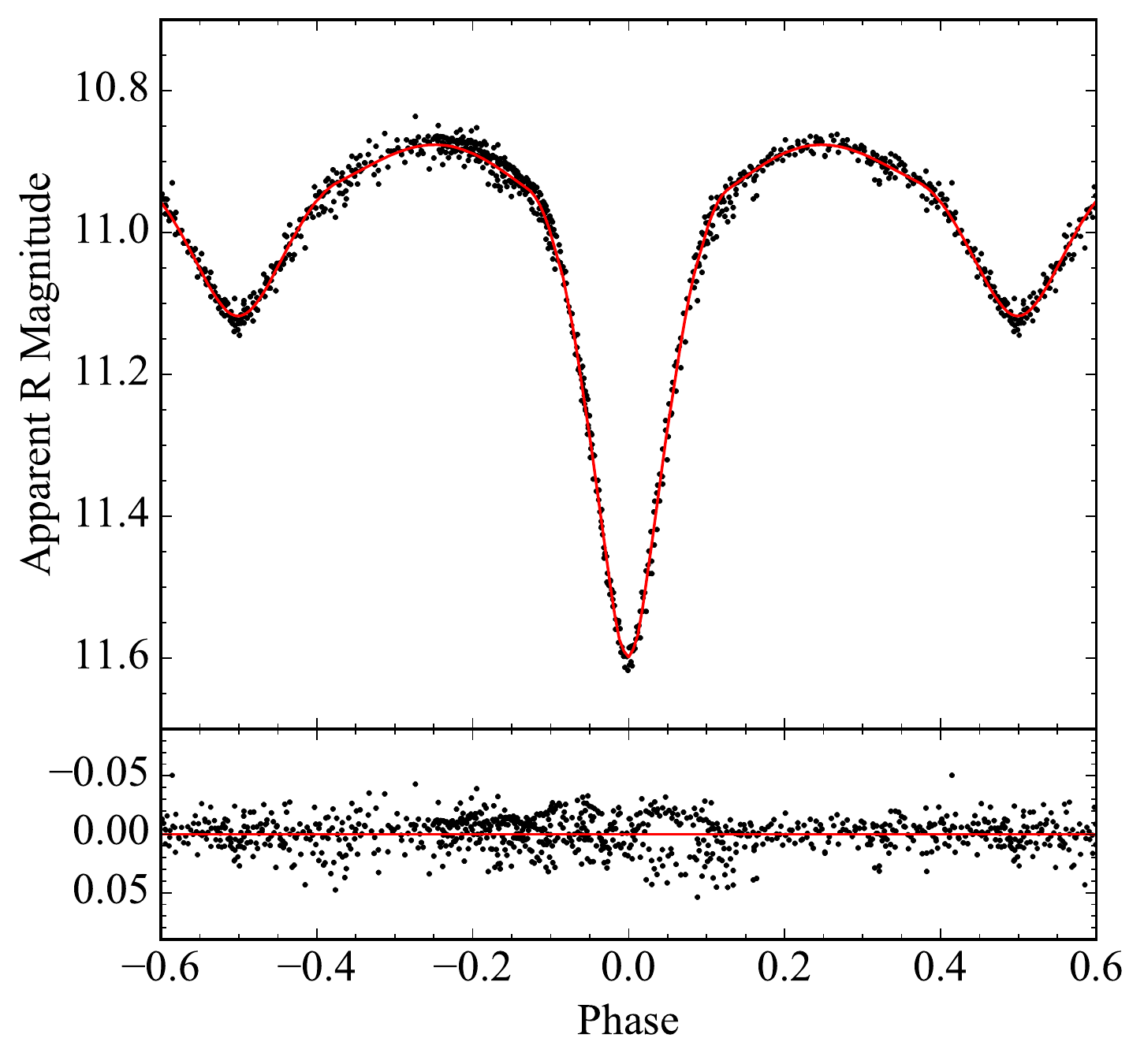}
\includegraphics[width=0.65\columnwidth]{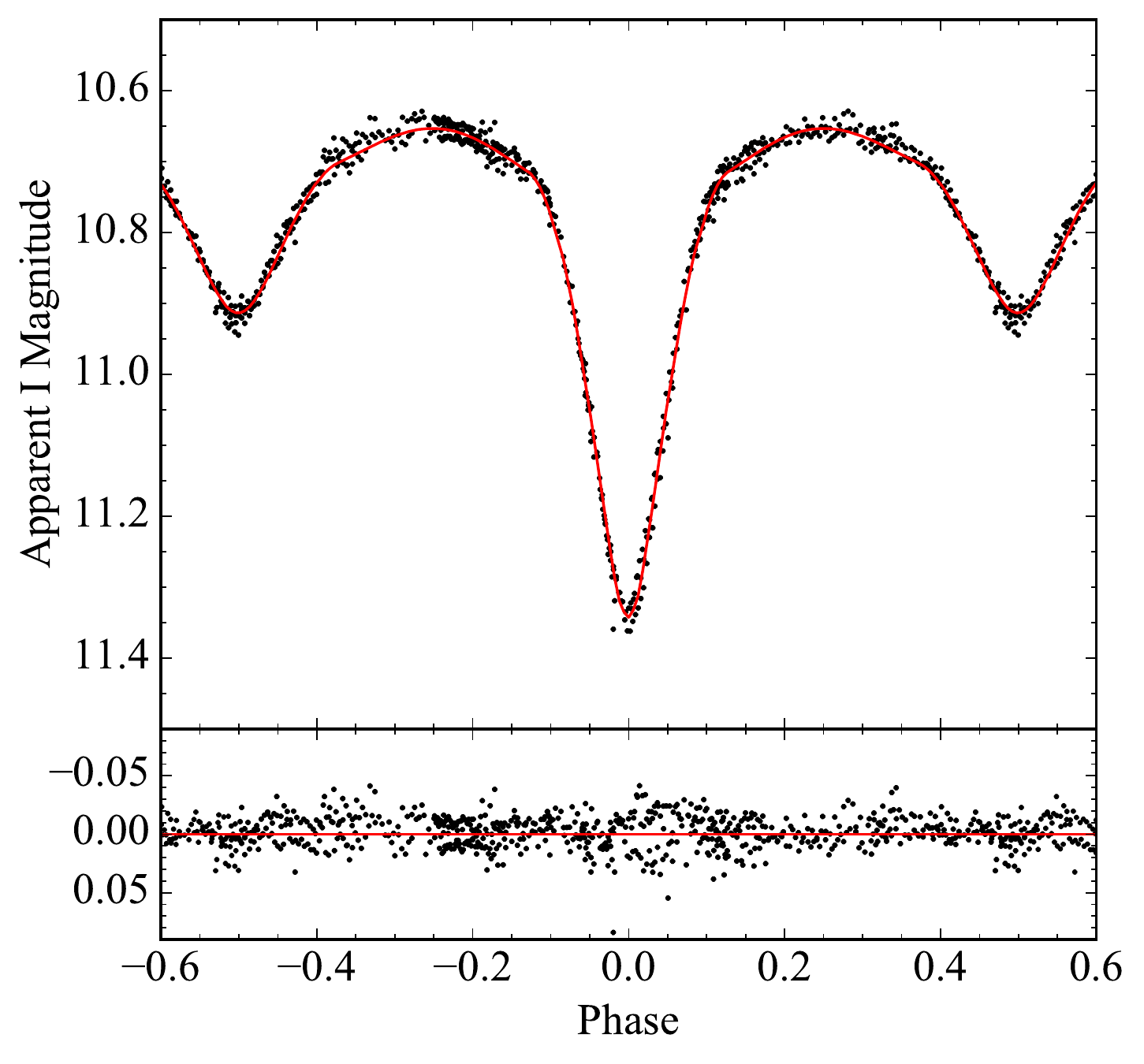}
\caption{Observed B (top panel), V (second panel), R (third panel), and I (fourth panel) data for NSVS 5726288 plotted against synthetic light curves. The top panel of each figure plots the apparent magnitude of the observed and synthetic light curves against the phase of the system. The residuals for the model are shown in the bottom panel of each figure.}
\label{fig:Tar-b-BVRI-models}
\end{center}
\end{figure}

A single, well-fitting model was produced for this system, the details of which can be found in the rightmost column of Table \ref{tab:Model-results}. The model indicates a detached system with a secondary component of spectral type G. It should be noted that, due to the lack of correction for interstellar extinction, the values for the temperature are a lower bound. A three-dimensional visualization of the model is given in Figure \ref{fig:Tar-b-3d-model} while a comparison of the observed data and synthetic light curves is presented in Figure \ref{fig:Tar-b-BVRI-models}.

\section{Discussion\label{sec:Discussion}}

NSVS 7322420 is a semi-detached system with the primary component filling its critical lobe. The system contains a G3 primary and an M secondary, and the model indicates that the stars are of significantly dissimilar mass. The light curve of the system displays a pronounced O'Connell effect and an unusual ``kink'' as the system enters and exits secondary eclipse. The system is also rapidly evolving, and data taken in 2013 and late 2014 could not be combined with data taken in early 2014 without producing obvious discontinuities that rendered modeling the system impossible. This light curve variability forced us to use only a subset of the data taken on the system that was relatively close temporally. The model produced is artificial and likely does not accurately explain these unusual features; a more sophisticated modeling program may be necessary.

NSVS 5726288 is a rather standard detached system containing two components of dissimilar mass in a moderately inclined orbit. The system contains an A8 primary and an early G secondary. The determined values for the temperature are lower bounds due to the lack of correcting for interstellar extinction, so it is possible that these stars are significantly hotter than our model indicates. The period of 0.856255 days differs significantly from the value published by \citet{2008AJ....136.1067H}, which is most likely a symptom of poor temporal coverage of the system by the NSVS.

The O'Connell effect observed in NSVS 7322420 is a poorly understood phenomenon. \citet{2009SASS...28..107W} describe several ideas regarding the cause of the effect, but they note that none of these ideas can adequately explain all known examples. The most prevalent explanation of the O'Connell effect is that hot or cool spots caused by chromospheric activity on one of the components creates a difference in the observed flux for different hemispheres of the star. However, the spots may need to be unrealistically large in size in order to fully describe the observed O'Connell effect, and spots also suffer from the ability to explain almost any deviation from a synthetic light curve if placed correctly, which reduces confidence in the reliability of such models. Furthermore, \citet{2014ApJS..213....9D} finds ``no evidence for changes in the maxima that are expected as star spot numbers or sizes vary,'' further diminishing the theory that star spots are the cause of the O'Connell effect. An alternate hypothesis proposed by \citet{2003ChJAA...3..142L} suggests that the effect is caused by material surrounding the stars impacting the components as they orbit, resulting in heating of the leading hemispheres. At this time, however, we lack the observational evidence to suggest a plausible source for such material in NSVS 7322420, and we consider it unlikely to explain the effect in the system.

Another hypothesis proposes that the asymmetry in observed flux is caused by a hot spot created by the impact of a matter stream on either the stellar surface or on a circumstellar accretion disk. This hypothesis has been used to explain the O'Connell effect observed in GR Tauri \citep{2004A&A...423..607G}, a system that exhibits light curve variability similar to what we observed in NSVS 7322420. Due to this and the previously mentioned irregularities in the light curve of the system (including an asymmetric primary minimum similar to RY Scuti, a system that was modeled with an accretion disk by \citealt{2008AJ....136..767D}), we believe that the mass transfer theory represents the most likely explanation of the O'Connell effect in NSVS 7322420.

Unfortunately, photometry alone cannot provide a full description of these systems. Radial velocity data obtained from spectra of these systems would allow us to determine the absolute masses of the components, and therefore the absolute sizes and luminosities of the stars. Spectroscopic data would also provide a direct way to measure the temperatures of the components, which would further refine the parameters of the models. Finally, spectroscopic analysis of NSVS 7322420 could provide clues as to the cause of the unusual features seen in its light curve. Further photometric and spectroscopic observations are being conducted on NSVS 732240 and will be described in a future paper. The system is also serving as an archetype for a study of systems suspected of undergoing mass transfer.

\section{Acknowledgements}

We would like to thank Drs. Eric Perlman and Saida Caballero-Nieves of FIT for their comments and suggestions regarding this paper. We would also like to thank the Indiana Space Grant Consortium, which partially funded this research project. This paper is based on observations obtained with the  SARA Observatory 0.9-meter telescope at Kitt Peak, which is owned and operated by the  Southeastern Association for Research in Astronomy (saraobservatory.org). The authors are honored to be permitted to conduct astronomical research on Iolkam Du'ag (Kitt Peak), a mountain with particular significance to the Tohono O'odham Nation.

\newpage

\bibliography{PaperDraftNotes}

\begin{thebibliography}{}
\expandafter\ifx\csname natexlab\endcsname\relax\def\natexlab#1{#1}\fi

\bibitem[{{Bessell}(1990)}]{1990PASP..102.1181B}
{Bessell}, M.~S. 1990, \pasp, 102, 1181

\bibitem[{{Bradstreet} \& {Steelman}(2002)}]{2002AAS...201.7502B}
{Bradstreet}, D.~H., \& {Steelman}, D.~P. 2002, in Bulletin of the American
  Astronomical Society, Vol.~34, American Astronomical Society Meeting
  Abstracts, 1224

\bibitem[{{Collins} {et~al.}(2017){Collins}, {Kielkopf}, {Stassun}, \&
  {Hessman}}]{2017AJ....153...77C}
{Collins}, K.~A., {Kielkopf}, J.~F., {Stassun}, K.~G., \& {Hessman}, F.~V.
  2017, \aj, 153, 77

\bibitem[{{Djura{\v s}evi{\'c}} {et~al.}(2008){Djura{\v s}evi{\'c}}, {Vince},
  \& {Atanackovi{\'c}}}]{2008AJ....136..767D}
{Djura{\v s}evi{\'c}}, G., {Vince}, I., \& {Atanackovi{\'c}}, O. 2008, \aj,
  136, 767

\bibitem[{{Drake} {et~al.}(2014){Drake}, {Graham}, {Djorgovski}, {Catelan},
  {Mahabal}, {Torrealba}, {Garc{\'{\i}}a-{\'A}lvarez}, {Donalek}, {Prieto},
  {Williams}, {Larson}, {Christen sen}, {Belokurov}, {Koposov}, {Beshore},
  {Boattini}, {Gibbs}, {Hill}, {Kowalski}, {Johnson}, \&
  {Shelly}}]{2014ApJS..213....9D}
{Drake}, A.~J., {Graham}, M.~J., {Djorgovski}, S.~G., {et~al.} 2014, \apjs,
  213, 9

\bibitem[{{Fitzgerald}(1970)}]{1970A&A.....4..234F}
{Fitzgerald}, M.~P. 1970, \aap, 4, 234

\bibitem[{{Flower}(1996)}]{1996ApJ...469..355F}
{Flower}, P.~J. 1996, \apj, 469, 355

\bibitem[{{Gu} {et~al.}(2004){Gu}, {Chen}, {Choy}, {Leung}, {Chung}, \&
  {Poon}}]{2004A&A...423..607G}
{Gu}, S.-h., {Chen}, P.-s., {Choy}, Y.-k., {et~al.} 2004, \aap, 423, 607

\bibitem[{{Henden} {et~al.}(2016){Henden}, {Templeton}, {Terrell}, {Smith},
  {Levine}, \& {Welch}}]{2016yCat.2336....0H}
{Henden}, A.~A., {Templeton}, M., {Terrell}, D., {et~al.} 2016, VizieR Online
  Data Catalog, II/336

\bibitem[{{Hoffman} {et~al.}(2008){Hoffman}, {Harrison}, {Coughlin},
  {McNamara}, {Holtzman}, {Taylor}, \& {Vestrand}}]{2008AJ....136.1067H}
{Hoffman}, D.~I., {Harrison}, T.~E., {Coughlin}, J.~L., {et~al.} 2008, \aj,
  136, 1067

\bibitem[{{Jester} {et~al.}(2005){Jester}, {Schneider}, {Richards}, {Green},
  {Schmidt}, {Hall}, {Strauss}, {Vanden Berk}, {Stoughton}, {Gunn},
  {Brinkmann}, {Kent}, {Smith}, {Tucker}, \& {Yanny}}]{2005AJ....130..873J}
{Jester}, S., {Schneider}, D.~P., {Richards}, G.~T., {et~al.} 2005, \aj, 130,
  873

\bibitem[{{Kafka}(2015)}]{stella_kafka_2015}
{Kafka}, S. 2015, {Observations from the AAVSO International Database}, https://www.aavso.org/aavso-international-database

\bibitem[{{Keel} {et~al.}(2017){Keel}, {Oswalt}, {Mack}, {Henson}, {Hillwig},
  {Batcheldor}, {Berrington}, {De Pree}, {Hartmann}, {Leake}, {Licandro},
  {Murphy}, {Webb}, \& {Wood}}]{2017PASP..129a5002K}
{Keel}, W.~C., {Oswalt}, T., {Mack}, P., {et~al.} 2017, \pasp, 129, 015002

\bibitem[{{Kwee} \& {van Woerden}(1956)}]{1956BAN....12..327K}
{Kwee}, K.~K., \& {van Woerden}, H. 1956, \bain, 12, 327

\bibitem[{{Liu} \& {Yang}(2003)}]{2003ChJAA...3..142L}
{Liu}, Q.-Y., \& {Yang}, Y.-L. 2003, \cjaa, 3, 142

\bibitem[{{Lucy}(1967)}]{1967ZA.....65...89L}
{Lucy}, L.~B. 1967, \zap, 65, 89

\bibitem[{{Milone}(1968)}]{1968AJ.....73..708M}
{Milone}, E.~E. 1968, \aj, 73, 708

\bibitem[{{O'Connell}(1951)}]{1951PRCO....2...85O}
{O'Connell}, D.~J.~K. 1951, Publications of the Riverview College Observatory,
  2, 85

\bibitem[{{Paunzen} \& {Vanmunster}(2016)}]{2016AN....337..239P}
{Paunzen}, E., \& {Vanmunster}, T. 2016, Astronomische Nachrichten, 337, 239

\bibitem[{{Pr{\v s}a} \& {Zwitter}(2005)}]{2005ApJ...628..426P}
{Pr{\v s}a}, A., \& {Zwitter}, T. 2005, \apj, 628, 426

\bibitem[{{Qian} {et~al.}(2007){Qian}, {Yuan}, {Soonthornthum}, {Zhu}, {He}, \&
  {Yang}}]{2007ApJ...671..811Q}
{Qian}, S.-B., {Yuan}, J.-Z., {Soonthornthum}, B., {et~al.} 2007, \apj, 671,
  811

\bibitem[{{Ruci{\'n}ski}(1969)}]{1969AcA....19..245R}
{Ruci{\'n}ski}, S.~M. 1969, \actaa, 19, 245

\bibitem[{{Schlafly} \& {Finkbeiner}(2011)}]{2011ApJ...737..103S}
{Schlafly}, E.~F., \& {Finkbeiner}, D.~P. 2011, \apj, 737, 103

\bibitem[{{Schwarzenberg-Czerny}(1996)}]{1996ApJ...460L.107S}
{Schwarzenberg-Czerny}, A. 1996, \apjl, 460, L107

\bibitem[{{Tody}(1993)}]{1993ASPC...52..173T}
{Tody}, D. 1993, in Astronomical Society of the Pacific Conference Series,
  Vol.~52, Astronomical Data Analysis Software and Systems II, ed. R.~J.
  {Hanisch}, R.~J.~V. {Brissenden}, \& J.~{Barnes}, 173

\bibitem[{{van Hamme}(1993)}]{1993AJ....106.2096V}
{van Hamme}, W. 1993, \aj, 106, 2096

\bibitem[{{Wilsey} \& {Beaky}(2009)}]{2009SASS...28..107W}
{Wilsey}, N.~J., \& {Beaky}, M.~M. 2009, Society for Astronomical Sciences
  Annual Symposium, 28, 107

\bibitem[{{Wilson} \& {Devinney}(1971)}]{1971ApJ...166..605W}
{Wilson}, R.~E., \& {Devinney}, E.~J. 1971, \apj, 166, 605

\bibitem[{{Wo{\'z}niak} {et~al.}(2004){Wo{\'z}niak}, {Vestrand}, {Akerlof},
  {Balsano}, {Bloch}, {Casperson}, {Fletcher}, {Gisler}, {Kehoe}, {Kinemuchi},
  {Lee}, {Marshall}, {McGowan}, {McKay}, {Rykoff}, {Smith}, {Szymanski}, \&
  {Wren}}]{2004AJ....127.2436W}
{Wo{\'z}niak}, P.~R., {Vestrand}, W.~T., {Akerlof}, C.~W., {et~al.} 2004, \aj,
  127, 2436

\end{thebibliography}

\end{document}